\documentclass[graybox]{svmult}

\usepackage{url}
\usepackage{geometry,
  , csquotes
  , enumitem
  , doi%
  , tikz %
  }
\usepackage{amssymb}
\usepackage{mathptmx}       
\usepackage{helvet}         
\usepackage{courier}        
\usepackage{type1cm}        

\usepackage{makeidx}         
\usepackage{graphicx}        
\usepackage{multicol}        
\usepackage[bottom]{footmisc}

\usepackage[english]{babel}
\usepackage[utf8x]{inputenc}
\usepackage[T1]{fontenc}
\usepackage{amsmath}
\usepackage{graphicx}
\usepackage[colorinlistoftodos]{todonotes}
\usepackage{mathptmx}      
\usepackage{helvet}        
\usepackage{courier}        
\usepackage{type1cm}                               
\usepackage{makeidx}        
\usepackage{graphicx}                                     
\usepackage{multicol}        
\usepackage[bottom]{footmisc}

\newcommand{\lto}[1]{\mathrel{\stackrel{{\;\;#1\;\;}}{\mbox{\rightarrowfill}}}}
\newcommand{\llto}[1]{\mathrel{\stackrel{{\;\;#1\;\;}}{\mbox{\leftarrowfill}}}}
\newcommand{\pnum}[1]{\raisebox{-1pt}{\large \textcircled{\raisebox{1.1pt} 
{\scriptsize #1}}}}

\makeindex 
\begin{document}

\title*{Microservices: yesterday, today, and tomorrow}
\authorrunning{N. Dragoni, S. Giallorenzo, A. Lluch~Lafuente, M. Mazzara et al.}
\author{Nicola Dragoni, Saverio Giallorenzo, Alberto Lluch Lafuente, Manuel Mazzara \\ Fabrizio Montesi, Ruslan Mustafin, Larisa Safina}
\institute{Nicola Dragoni, Alberto Lluch Lafuente \at Technical University of Denmark \email{{ndra, albl}@dtu.dk}
\and Saverio Giallorenzo\at INRIA/Univ. of Bologna, Dept. of Computer Science and Engineering, Italy \email{saverio.giallorenzo@gmail.com}
\and  Manuel Mazzara, Ruslan Mustafin, Larisa Safina \at Innopolis University, Russian Federation \email{{ m.mazzara, r.mustafin, l.safina}@innopolis.ru}
\and Fabrizio Montesi \at University of Southern Denmark
\email{fmontesi@imada.sdu.dk}
}

\maketitle
\abstract{Microservices is an architectural style inspired by service-oriented computing that has recently started gaining popularity. Before presenting the current state-of-the-art in the field, this chapter reviews the history of software architecture, the reasons that led to the diffusion of objects and services first, and microservices later. Finally, open problems and future challenges are introduced. This survey primarily addresses newcomers to the discipline, while offering an academic viewpoint on the topic. In addition, we investigate some practical issues and point out some potential solutions.}

\section{Introduction} 
\label{sec:introduction}
The mainstream languages for development of server-side applications, like Java, C/C++, and Python, provide abstractions to break down the complexity of
programs into modules.
However, these languages are designed for the creation of \emph{single
executable artefacts}, also called \emph{monoliths}, and their modularisation
abstractions rely on the sharing of resources of the same machine (memory, databases, files).
Since the modules of a monolith depend on said shared resources, they are not independently executable.

\begin{definition}[Monolith]
A monolith is a software application whose modules cannot be executed independently.
\end{definition}

This makes monoliths difficult to use in distributed systems
without specific frameworks or ad hoc solutions such as,
for example, Network Objects \cite{Birrell:1993}, RMI \cite{Grosso:2001} or
CORBA \cite{CORBA}.
However, even these approaches still suffer from the general issues that affect monoliths; below we list the most relevant ones (we label issues \textbf{I}$|n$):

\begin{enumerate}[label=\textbf{I}$|$\arabic*]

  \item large-size monoliths are difficult to maintain and evolve due to
  their complexity. Tracking down bugs requires long perusals through their
  code base;
  
  \item monoliths also suffer from the ``dependency
  hell''~\cite{dependency_hell}, in which adding or updating libraries results
  in inconsistent systems that do not compile/run or, worse, misbehave;
  
  \item any change in one module of a monolith requires rebooting the whole
  application. For large-sized projects, restarting usually entails considerable
  downtimes, hindering development, testing, and the maintenance of the
  project;

  \item deployment of monolithic applications is usually sub-optimal due to
  conflicting requirements on the constituent models' resources: some can
  be memory-intensive, others computational-intensive, and others require
  ad-hoc components (e.g., SQL-based rather than graph-based databases). When
  choosing a deployment environment, the developer must compromise with a
  one-size-fits-all configuration, which is either expensive or sub-optimal
  with respect to the individual modules;

  \item monoliths limit scalability. The usual strategy for handling increments
  of inbound requests is to create new instances of the same application and
  to split the load among said instances. However, it could be the case that
  the increased traffic stresses only a subset of the modules, making the
  allocation of the new resources for the other components inconvenient;

  \item monoliths also represent a technology lock-in for developers, which
  are bound to use the same language and frameworks of the original
  application.

\end{enumerate}

\vspace{1em}

The \emph{microservices} architectural style~\cite{original_article} has been
proposed to cope with such problems. In our definition of microservice, we use the term
``cohesive''~\cite{Dhama95,hitz95,Bieman95,Briand99,Allen01} to indicate that
a service implements only functionalities strongly related to the concern that it
is meant to model.

\begin{definition}[Microservice] A microservice is a cohesive, independent
process interacting via messages.
\end{definition}

As an example, consider a service intended to compute
calculations. To call it a microservice, it should provide arithmetic operations 
requestable via messages,  but it should not provide other (possibly loosely related)
functionalities like plotting and displaying of functions.

From a technical point of view, microservices should be independent components
conceptually deployed in isolation and equipped with dedicated memory
persistence tools (e.g., databases). Since all the components of a
microservice architecture are microservices, its distinguishing behaviour
derives from the composition and coordination of its components via messages.

\begin{definition}[Microservice Architecture] A microservice architecture is a
distributed application where all its modules are microservices.
\end{definition}

To give an example of a microservice architecture, let us assume that we want
to provide a functionality that plots the graph of a function. We also assume
the presence of two microservices: Calculator and Displayer. The first is the
calculator microservice mentioned above, the second renders and displays
images. To fulfil our goal, we can introduce a new microservice, called
Plotter, that orchestrates Calculator to calculate the shape of the graph and
that invokes Displayer to render the calculated shape. Below, we report (in
black) a depiction of the workflow of such a microservice architecture.
$$\begin{array}{ccccccccc}
\mbox{user input} & \lto{\ \pnum{1}\ } & 
\fbox{Plotter} & \lto{\ \pnum{2}\ } & \fbox{\hspace{1em} Calculator \hspace{1em}}
& {\color{gray}\longrightarrow} & {\color{gray}\fbox{Elementary Functions}}
\\[.5em] && \downarrow \pnum{3} && {\color{gray} \downarrow}\\
\mbox{user display} & \llto{\ \pnum{4}\ } & \fbox{Displayer} &&
{\color{gray}\fbox{Special Functions}}
\end{array}$$
The developers of the architecture above can focus separately on
implementing the basic microservice functionalities, i.e., the
Calculator and the Displayer. Finally, they can implement the behaviour of the
distributed application with the Plotter that \pnum{1} takes the function
given by a user, \pnum{2} interacts with the Calculator to compute a symbolic
representation of the graph of the function, and finally  \pnum{3} requests
the Displayer to show the result back to the user \pnum{4}. To illustrate how
the microservice approach scales by building on pre-existing microservice
architectures, in the figure above we drew the Calculator orchestrating two
extra microservices (in grey) that implement mathematical Elementary and
Special Functions.

The microservice architectural style does not favour or forbid any particular
programming paradigm. It provides a guideline to partition the components of a
distributed application into independent entities, each addressing one of its
concerns. This means that a microservice, provided it 
offers its functionalities via message passing, can be internally implemented 
with any of the mainstream languages cited in the beginning of this section.

The principle of microservice architectures assists project managers and
developers: it provides a guideline for the design and implementation of
distributed applications. Following this principle, developers focus on the
implementation and testing of a few, cohesive functionalities. This holds also
for higher-level microservices, which are concerned with coordinating the
functionalities of other microservices.



We conclude this section with an overview, detailed in greater depth in the
remainder of the paper, on how microservices cope with the mentioned issues
of monolithic applications (below, $\textbf{S}|n$ is a solution to issue
$\textbf{I}|n$).

\begin{enumerate}[label=\textbf{S}$|$\arabic*]

  \item microservices implement a limited amount of functionalities, which
  makes their code base small and inherently limits the scope of a bug.
  Moreover, since microservices are independent, a developer can directly test
  and investigate their functionalities in isolation with respect to the rest
  of the system;
  
  \item it is possible to plan \emph{gradual transitions} to new versions of a
  microservice. The new version can be deployed ``next'' to the old one and
  the services that depend on the latter can be gradually modified to interact
  with the former. This fosters \emph{continuous
  integration}~\cite{fowler2006continuous} and greatly eases software
  maintenance;
  
  \item as a consequence of the previous item, changing a module of a
  microservice architecture does not require a complete reboot of the whole
  system. The reboot regards only the microservices of that module. Since 
  microservices  are small in size, programmers can develop, test, and 
  maintain services experiencing only very short re-deployment downtimes;

  \item microservices naturally lend themselves to
  containerisation~\cite{docker}, and developers enjoy a high degree of
  freedom in the configuration of the deployment environment that best suits
  their needs (both in terms of costs and quality of service);

  \item scaling a microservice architecture does not imply a duplication of
  all its components and developers can conveniently deploy/dispose instances
  of services with respect to their load~\cite{GabbrielliGGMM16};

  \item the only constraint imposed on a network of interoperating
  microservices is the technology used to make them communicate (media,
  protocols, data encodings). Apart from that, microservices impose no
  additional lock-in and developers can freely choose the optimal resources
  (languages, frameworks, etc.) for the implementation of
  each microservice.

\end{enumerate}

\vspace{1em}

In the remainder of this paper, in \S~\ref{sec:yesterday}, we give a brief
account of the evolution of distributed architectures until their recent
incarnation in the microservice paradigm. Then, we detail the problems that
microservices can solve and their proposed solutions in the form of
microservice architectures. In \S~\ref{sec:today}, we detail the current
solutions for developing microservice architectures and how microservices
affect the process of software design, development, testing, and maintenance.
In \S~\ref{sec:tomorrow} we discuss the open challenges and the desirable
tools for programming microservice architecture. In \S~\ref{sec:conclusions}
we draw overall conclusions.



\section{Yesterday}
\label{sec:yesterday}
Architecture is what allows systems to evolve and provide a certain level of service throughout their life-cycle. In software engineering, architecture is concerned with providing a bridge between system functionality and requirements for quality attributes that the system has to meet. Over the past several decades, software architecture has been thoroughly studied, and as a result software engineers have come up with different ways to compose systems that provide broad functionality and satisfy a wide range of requirements. In this section, we provide an overview of the work on software architectures from the early days to the advent of microservices.
 
\subsection{From the early days to Object-oriented design patterns}

The problems associated with large-scale software development were first experienced around the 1960s~\cite{brooks1975mythical}. The 1970s saw a huge rise of interest from the research community for software design and its implications on the development process. At the time, the design was often considered as an activity not associated with the implementation itself and therefore requiring a special set of notations and tools. Around the 1980s, the full integration of design into the development processes contributed towards a partial merge of these two activities, thus making it harder to make neat distinctions.

References to the concept of software architecture also started to appear around the 1980s. However, a solid foundation on the topic was only established in 1992 by Perry and Wolf~\cite{perry1992foundations}. Their definition of software architecture was distinct from software design, and since then it has generated a large community of researchers studying the notion and the practical applications of software architecture, allowing the concepts to be widely adopted by both industry and academia.

This spike of interest contributed to an increase in the number of existing software architecture patterns (or generally called \textit{styles}), so that some form of classification was then required. This problem was tackled in one of the most notable works in the field, the book ``Software Architecture: Perspectives on an Emerging Discipline'' by Garlan and Shaw~\cite{shaw1996software}. Bosch's work ~\cite{bosch2004software} provides a good overview of the current research state in software engineering and architecture. Since its appearance in the 1980s, software architecture has developed into a mature discipline making use of notations, tools, and several techniques. From the pure, and occasionally speculative, realm of academic basic research, it has made the transition into an element that is essential to industrial software construction. 

The advent and diffusion of object-orientation, starting from the 1980s and in particular in the 1990s, brought its own contribution to the field of Software Architecture. The classic by Gamma et al.~\cite{gamma1995design} covers the design of object-oriented software and how to translate it into code presenting a collection of recurring solutions, called \emph{patterns}. This idea is neither new nor exclusive to Software Engineering, but the book is the first compendium to popularize the idea on a large scale. In the pre-Gamma era patterns for OO solutions were already used: a typical example of an architectural design pattern in object-oriented programming is the Model-View-Controller (MVC)~\cite{Fowler2002}, which has been one of the seminal insights in the early development of graphical user interfaces.

\subsection{Service-oriented Computing}

Attention to \emph{separation of concerns} has recently led to the emergence of the so-called Component-based software engineering (CBSE)~\cite{Szyperski:2002}, which has given better control over design, implementation and evolution of software systems. The last decade has seen a further shift towards the concept of service first~\cite{ws:website} and the natural evolution to microservices afterwards. 

Service-Oriented Computing (SOC) is an emerging paradigm for distributed computing and e-business processing that finds its origin in object-oriented and component computing. It has been introduced to harness the complexity of distributed systems and to integrate different software applications~\cite{mackenzie2006}. In SOC, a program --- called a \emph{service} --- offers functionalities to other components, accessible via message passing. Services decouple their interfaces (i.e. how other services access their functionalities) from their implementation. On top of that, specific workflow languages are then defined in order to orchestrate the complex actions of services (e.g. WS-BPEL~\cite{bpel}).
These languages share ideas with some well-known formalisms from concurrency theory, such as CCS and the $\pi$-calculus~\cite{M80,MPW92}.
This aspect fostered the development of formal models for better understanding and verifying service interactions, ranging from foundational process models of SOC~\cite{Mazzara:phd,Guidi:phd,LucchiM07} to theories for the correct composition of services~\cite{bravetti2007towards,HVK98,HYC16}. In \cite{YanMCU07} a classification of approaches for business modeling puts this research in perspective.

The benefits of service-orientation are:
\\
\begin{itemize}
    \item \textbf{Dynamism} - New instances of the same service can be launched to split the load on the system;
    \item \textbf{Modularity and reuse} - Complex services are composed of simpler ones. The same services can be used by different systems;
    \item \textbf{Distributed development} - By agreeing on the interfaces of the distributed system, distinct development teams can develop partitions of it in parallel;
    \item \textbf{Integration of heterogeneous and legacy systems} - Services merely have to implement standard protocols to communicate.
\end{itemize}

\subsection{Second generation of services}

The idea of componentization used in service-orientation can be partially traced back to the object-oriented programming (OOP) literature; however, there are peculiar differences that led to virtually separate research paths and communities. As a matter of fact, SOC at the origin was - and still is - built on top of OOP languages, largely due to their broad diffusion in the early 2000s. However, the evolution of objects into services, and the relative comparisons, has to be treated carefully since the first focus on encapsulation and information is hidden in a \textit{shared-memory} scenario, while the second is built on the idea of independent deployment and \textit{message-passing}. It is therefore a paradigm shift, where both the paradigms share the common idea of componentization. The next step is adding the notion of \textit{business capability} and therefore focusing analysis and design on it so that  the overall system architecture is determined on this basis.

The first ``generation'' of service-oriented architectures (SOA) defined daunting and nebulous requirements for services (e.g., discoverability and service contracts), and this hindered the adoption of the SOA model. Microservices are the second iteration on the concept of SOA and SOC. The aim is to strip away unnecessary levels of complexity in order to focus on the programming of simple services that effectively implement a single functionality. Like OO, the microservices paradigm needs ad-hoc tools to support developers and naturally leads to the emergence of specific design patterns~\cite{Safina2016}. First and foremost, languages that embrace the service-oriented paradigm are needed (instead, for the most part, microservice architectures still use OO languages like Java and Javascript or functional ones). The same holds for the other tools for development support like testing suites, (API) design tools, etc.

\section{Today}
\label{sec:today}
The microservices architecture appeared lately as a new paradigm for programming applications by means of the composition of small services, each running its own processes and communicating via light-weight mechanisms. This approach has been built on the concepts of SOA~\cite{mackenzie2006} brought from crossing-boundaries workflows to the application level and into the applications architectures, i.e. its Service-Oriented Architecture and Programming from the large to the small. 

The term ``microservices'' was first introduced in 2011 at an architectural workshop as a way to describe the participants' common ideas in software architecture patterns~\cite{original_article}. Until then, this approach had also been known under different names. For example, Netflix used a very similar architecture under the name of \emph{Fine grained SOA}~\cite{wang2013ribbon}. 

Microservices now are a new trend in software architecture, which emphasises the design and development of highly maintainable and scalable software. Microservices manage growing complexity by functionally decomposing large systems into a set of independent services. By making services completely independent in development and deployment, microservices emphasise loose coupling and high cohesion by taking modularity to the next level. This approach delivers all sorts of benefits in terms of maintainability, scalability and so on. It also comes with a bundle of problems that are inherited from distributed systems and from SOA, its predecessor. The Microservices architecture still shows distinctive characteristics that blend into something unique and different from SOA itself:
\\
\begin{itemize}
\item \textbf{Size} - The size is comparatively small wrt. a typical service, supporting the belief that the architectural design of a system is highly dependent on the structural design of the organization producing it. Idiomatic use of the  microservices architecture suggests that if a service is too large, it should be split into two or more services, thus preserving granularity and maintaining focus on providing only a single business capability. This brings benefits in terms of service maintainability and extendability.

\item \textbf{Bounded context} - Related functionalities are combined into a single business capability, which is then implemented as a service.

\item \textbf{Independency} - Each service in microservice architecture is operationally independent from other services and the only form of communication between services is through their published interfaces.
\\
\end{itemize}

The key system characteristics for microservices are:
\\
\begin{itemize}
\item \textbf{Flexibility} - A system is able to keep up with the ever-changing business environment and is able to support all modifications that is necessary for an organisation to stay competitive on the market

\item \textbf{Modularity} - A system is composed of isolated components where each component contributes to the overall system behaviour rather than having a single component that offers full functionality

\item \textbf{Evolution} - A system should stay maintainable while constantly evolving and adding new features
\\
\end{itemize}

The microservices architecture gained popularity relatively recently and can be considered to be in its infancy since there is still a lack of consensus on what microservices actually are. M. Fowler and J. Lewis provide a starting ground by defining principal characteristics of microservices~\cite{original_article}. S. Newman~\cite{newman2015building} builds upon M. Fowler's article and presents recipes and best practices regarding some aspects of the aforementioned architecture. L. Krause in his work~\cite{krause2014microservices} discusses patterns and applications of microservices.  A number of papers has also been published that describe details of design and implementation of systems using microservices architecture. For example, the authors of~\cite{lemicroservice} present development details of a new software system for Nevada Research Data Center (NRDC) using the microservices architecture.  M. Rahman and J. Gao in~\cite{gao1reusable} describe an application of behaviour-driven development (BDD) to the microservices architecture in order to decrease the maintenance burden on developers and encourage the usage of acceptance testing. 

\subsection{Teams}
Back in 1968, Melvin Conway proposed that an organisation's structure, or more specifically, its communication structure constrains a system's design such that the resulting design is a copy of the organisation's communication patterns~\cite{conway1968committees}. The microservices approach is to organise cross-functional teams around services, which in turn are organised around business capabilities~\cite{original_article}. This approach is also known as ``you build, you run it'' principle, first introduced by Amazon CTO Werner Vogels~\cite{gray2006conversation}. According to this approach, teams are responsible for full support and development of a service throughout its lifecycle.

\subsection{Total automation}

Each microservice may represent a single business capability that is delivered and updated independently and on its own schedule. Discovering a bug and or adding a minor improvement do not have any impact on other services and on their release schedule (of course, as long as backwards compatibility is preserved and a service interface remains unchanged). However, to truly harness the power of independent deployment, one must utilise very efficient integration and delivery mechanisms. This being said, microservices are the first architecture developed in the post-continuous delivery era and essentially microservices are meant to be used with continuous delivery and continuous integration, making each stage of delivery pipeline automatic. By using automated continuous delivery pipelines and modern container tools, it is possible to deploy an updated version of a service to production in a matter of seconds~\cite{nginx2015teams}, which proves to be very beneficial in rapidly changing business environments.

\subsection{Choreography over orchestration}
As discussed earlier, microservices may cooperate in order to provide more complex and elaborate functionalities. There are two approaches to establish this cooperation -- orchestration~\cite{Mazzara2005} and choreography~\cite{P03}. Orchestration requires a conductor -- a central service that will send requests to other services and oversee the process by receiving responses. Choreography, on the other hand, assumes no centralisation and uses events and publish/subscribe mechanisms in order to establish collaboration. These two concepts are not new to microservices, but rather are inherited from the SOA world where languages such as WS-BPEL~\cite{bpel} and WS-CDL~\cite{WS-CDL} have long represented the major references for orchestration and choreography respectively (with vivid discussions between the two communities of supporters). 

Prior to the advent of microservices and at the beginning of the SOA's hype in particular, orchestration was generally more popular and widely adopted, due to its simplicity of use and easier ways to manage complexity. However, it clearly leads to 
service coupling and uneven distribution of responsibilities, and therefore some services have a more centralising role than others.
Microservices' culture of decentralisation and the high degrees of independence represents instead the natural application scenario
for the use of choreography as a means of achieving collaboration. 
This approach has indeed recently seen a renewed interest in connection with the broader diffusion of microservices in what can be called the\
\emph{second wave of services}.

\subsection{Impact on quality and management} 
In order to better grasp microservices we need to understand the impact that this architecture has on some software quality attributes. 

\paragraph{\textbf{Availability}}
\label{subsec:availability}
Availability is a major concern in microservices as it directly affects the success of a system. Given services independence, the whole system availability can be estimated in terms of the availability of the individual services that compose the system. Even if a single service is not available to satisfy a request, the whole system may be compromised and experience direct consequences. 
If we take service implementation, the more fault-prone a component is, the more frequently the system will experience failures. One would argue that small-size components lead to a lower fault density. However, it has been found by Hatton~\cite{hatton1997reexamining}  and by Compton and Withrow~\cite{compton1990prediction} that small-size software components often have a very high fault density. On the other hand, El Emam \textit{et al.} in their work~\cite{el2002optimal} found that as size increases, so does a component's fault proneness. Microservices are prevented from becoming too large as idiomatic use of the microservices architecture suggests that, as a system grows larger, microservices should be prevented from becoming overly complex by refining them into two or more different services. Thus, it is possible to keep optimal size for services, which may theoretically increase availability. On the other hand, spawning an increasing number of services will make the system fault-prone on the integration level, which will result in decreased availability due to the large complexity associated with making dozens of services instantly available.

\paragraph{\textbf{Reliability}}
Given the distributed nature of the microservices architecture, particular attention should be paid to the reliability of message-passing mechanisms between services and to the reliability of the services themselves. 
Building the system out of small and simple components is also one of the rules introduced in~\cite{raymond2003art}, which states that in order to achieve higher reliability one must find a way to manage the complexities of a large system: building things out of simple components with clean interfaces is one way to achieve this. The greatest threat to microservices reliability lies in the domain of integration and therefore when talking about microservices reliability, one should also mention integration mechanisms. One example of this assumption being false is using a network as an integration mechanism and assuming network reliability is one of the first fallacies of distributed computing~\cite{rotem2006fallacies}. Therefore, in this aspect, microservices reliability is inferior to the applications that use in-memory calls. It should be noted that this downside is not unique only to microservices and can be found in any distributed system.
When talking about messaging reliability, it is also useful to remember that microservices put restrictions on integration mechanisms. More specifically, microservices use integration mechanisms in a very straightforward way - by removing all functionality that is not related to the message delivering and focusing solely on reliable message delivery. 

\paragraph{\textbf{Maintainability}}
By nature, the microservices architecture is loosely coupled, meaning that there is a small number of links between services and services themselves being independent. This greatly contributes to the maintainability of a system by minimising the costs of modifying services, fixing errors or adding new functionality.
Despite all efforts to make a system as maintainable as possible, it is always possible to spoil maintainability by writing obscure and counterintuitive code~\cite{bass2007software}. As such, another aspect of microservices that can lead to increased maintainability is the above mentioned ``you build it, you run it'' principle, which leads to better understanding a given service, its business capabilities and roles~\cite{fagan2002design,cohen2006best}. 

\paragraph{\textbf{Performance}}
The prominent factor that negatively impacts performance in the microservices architecture is communication over a network. The network latency is much greater than that of memory. This means that in-memory calls are much faster to complete than sending messages over the network. Therefore, in terms of communication, the performance will degrade compared to applications that use in-memory call mechanisms. Restrictions that microservices put on size also indirectly contribute to this factor. In more general architectures without size-related restrictions, the ratio of in-memory calls to the total number of calls is higher than in the microservices architecture, which results in less communication over the network.  Thus, the exact amount of performance degradation will also depend on the system's interconnectedness. As such, systems with well-bounded contexts will experience less degradation due to looser coupling and fewer messages sent. 

\paragraph{\textbf{Security}}
In any distributed system security becomes a major concern. In this sense, microservices suffer from the same security vulnerabilities as SOA~\cite{bass2005quality}. As microservices use REST mechanism and XML with JSON as main data-interchange formats, particular attention should be paid to providing security of the data being transferred. This means adding additional overhead to the system in terms of additional encryption functionality. Microservices promote service reuse, and as such it is natural to assume that some systems will include third-party services. Therefore, an additional challenge is to provide authentication mechanisms with third-party services and ensure that the sent data is stored securely. In summary, microservices' security is impacted in a rather negative manner because one has to consider and implement additional security mechanisms to provide additional security functionality mentioned above. 

\paragraph{\textbf{Testability}}
Since all components in a microservices architecture are independent, each component can be tested in isolation, which significantly improves component testability compared to monolithic architecture. It also allows to adjust the scope of testing based on the size of changes. This means that with microservices it is possible to isolate parts of the system that changed and parts that were affected by the change and to test them independently from the rest of the system. Integration testing, on the other hand, can become very tricky, especially when the system that is being tested is very large, and there are too many connections between components. It is possible to test each service individually, but anomalies can emerge from collaboration of a number of services. 


\section{Tomorrow}
\label{sec:tomorrow}
Microservices are so recent that we can consider their exploration to have just begun. In this section, we discuss interesting future directions that we envision will play key roles in the advancement of the paradigm.

The greatest strength of microservices comes from pervasive distribution: even
the internal components of software are autonomous services, leading to
loosely coupled systems and the other benefits previously discussed. However,
from this same aspect (distribution) also comes its greatest weakness:
programming distributed systems is inherently harder than monoliths. We now
have to think about new issues. Some examples are: how can we manage changes to a
service that may have side-effects on the other services that it communicates
with? How can we prevent attacks that exploit network communications?

\subsection{Dependability}

There are many pitfalls that we need to keep in mind when programming with
microservices. In particular, preventing programming errors is hard. Consequently, building dependable systems is challenging.

\paragraph{\textbf{Interfaces}} Since microservices are autonomous, we are free to use
the most appropriate technology for the development of each microservice. A
disadvantage introduced by this practice is that different technologies
typically have different means of specifying contracts for the composition of
services (e.g., interfaces in Java, or WSDL documents in Web
Services~\cite{wsdl}). Some technologies do not even come with a specification
language and/or a compatibility checker of microservices (Node.js, based on
JavaScript, is a prime example).

Thus, where do we stand? Unfortunately, the current answer is informal
documentation. Most services come with informal documents expressed in natural
language that describe how clients should use the service. This makes the
activity of writing a client very error-prone, due to potential ambiguities.
Moreover, we have no development support tools to check whether service
implementations actually implement their interfaces correctly.

As an attempt to fix this problem, there are tools for the formal
specification of message types for data exchange, which one can use to define
service interfaces independently of specific technologies. Then, these
technology-agnostic specifications can be either compiled to language-specific
interfaces --- e.g., compiling an interface to a Java type --- or used to
check for well-typedness of messages (wrt. interfaces and independently of the
transport protocol). Examples of tools offering these methodologies are
Jolie~\cite{MGZ14,Bandura16,Guidi2017}, Apache Thrift~\cite{thrift}, and Google's Protocol
Buffers~\cite{protobuf}. However, it is still unclear how to adapt tools to
implement the mechanical checking (at compile or execution time) of messages
for some widespread architectural styles for microservices, such as
REST~\cite{F00}, where interfaces are constrained to a fixed set of operations
and actions are expressed on dynamic resource paths. A first attempt at
bridging the world of technology-agnostic interfaces based on operations and
REST is presented in~\cite{M16}, but checking for the correctness of the
binding information between the two is still left as a manual task to the
programmer. Another, and similar, problem is trying to apply static type checking to dynamic languages (e.g., JavaScript and Jolie), which are largely employed in the development of microservices~\cite{flow:website,mingela2017,akentev2017}.

\paragraph{\textbf{Behavioural Specifications and Choreographies}} Having formally-defined interfaces in the form of an API is not enough to guarantee the compatibility of services. This is because, during execution, services may
engage in sessions during which they perform message exchanges in a precise
order. If two services engage in a session and start performing incompatible
I/O, this can lead to various problems. Examples include: a client sending a
message on a stream that was previously closed; deadlocks, when two services
expect a message from one another without sending anything; or, a client
trying to access an operation that is offered by a server only after a
successful distributed authentication protocol with a third-party is
performed.

Behavioural types are types that can describe the behaviour of services and
can be used to check that two (or more) services have compatible actions.
Session types are a prime example of behavioural types~\cite{HVK98,HYC16}.
Session types have been successfully applied to many contexts already, ranging
from parallel to distributed computing. However, no behavioural type theory is
widely adopted in practice yet. This is mainly because behavioural types
restrict the kind of behaviours that programmers can write for services,
limiting their applicability. An important example of a feature with space for
improvement is non-determinism. In many interesting protocols, like those for
distributed agreement, execution is non-deterministic and depending on what
happens at runtime, the participants have to react differently~\cite{OO14}.

Behavioural interfaces are a hot topic right now and will likely play an
important role in the future of microservices. We envision that they will also
be useful for the development of automatic testing frameworks that check the
communication behaviour of services.

\paragraph{\textbf{Choreographies}} Choreographies are high-level descriptions of the
communications that we want to happen in a system in contrast with the
typical methodology of defining the behaviour of each service separately.
Choreographies are used in some models for behavioural interfaces, but they
actually originate from efforts at the W3C of defining a language that
describes the global behaviour of service systems~\cite{wscdl}. Over the past
decade, choreographies have been investigated for supporting a new programming
paradigm called Choreographic Programming~\cite{M13:phd}. In Choreographic
Programming, the programmer uses choreographies to program service systems and
then a compiler is used to automatically generate compliant implementations.
This yields a correctness-by-construction methodology, guaranteeing important
properties such as deadlock-freedom and lack of communication
errors~\cite{CHY12,CM13,MY13}.

Choreographies may have an important role in the future of microservices,
since they shrink the gap between requirements and implementations, making the
programmer able to formalise the communications envisioned in the design phase
of software. Since the correctness of the compiler from choreographies to
distributed implementations is vital in this methodology, formal models are
being heavily adopted to develop correct compilation algorithms~\cite{GGM15}.
However, a formalisation of how transparent mobility of processes from one
protocol to the other is still missing. Moreover, it is still unclear how
choreographies can be combined with flexible deployment models where nodes may
be replicated or fail at runtime. An initial investigation on the latter is
given in~\cite{LNN16}. Also, choreographies are still somewhat limited in
expressing non-deterministic behaviour, just like behavioural types.

\paragraph{\textbf{Moving Fast with Solid Foundations}} Behavioural types, choreographies, refinement types \cite{Tchitchigin16} and other models address the problem of specifying, verifying, and synthesising communication behaviours. However, there is still much to be discovered and developed on these topics. It is then natural to ask: do we
really need to start these investigations from scratch? Or, can we hope to
reuse results and structures from other well-established models in Computer
Science?

A recent line of work suggests that a positive answer can be
found by connecting behavioural types and choreographies to well-known
logical models. A prominent example is a Curry-Howard correspondence between
session types and the process model of $\pi$-calculus, given in~\cite{CP10}
(linear logical propositions correspond to session types, and communications
to proof normalization in linear logic). This result has propelled many other
results, among which:
a logical reconstruction of behavioural types in classical linear logic that supports parametric polymorphism~\cite{W14};
type theories for integrating higher-order process models with functional computation~\cite{TCP13};
initial ideas for
algorithms for extracting choreographies from separate service
programs~\cite{CMS14}; a logical characterisation of choreography-based behavioural types~\cite{CMSY15}; and, explanations of how interactions among multiple services (multiparty sessions) are related to well-known techniques for logical reasoning~\cite{CLMSW16,CP16}.

Another principle that we can use for the evolution of choreographic models is
the established notion of computation. The minimal set of language features to
achieve Turing completeness in choreographies is known~\cite{CM15}. More
relevant in practice, this model was used to develop a methodology of
procedural programming for choreographies, allowing for the writing of
correct-by-construction implementations of divide-and-conquer distributed
algorithms~\cite{CM16:divided}.

We can then conclude that formal methods based on well-known techniques seem to be a promising starting point for tackling the issue of writing correct microservice systems. This starting point gives us solid footing for exploring the more focused disciplines that we will need in the future, addressing problems like the description of coordination patterns among services. We envision that these patterns will benefit from the rich set of features that formal languages and process models have to offer, such as expressive type theories and logics. It is still unclear, however, how exactly these disciplines can be extended to naturally capture the practical scenarios that we encounter in microservices. We believe that empirically investigating microservice programming will be beneficial in finding precise research directions in this regard. 

\subsection{Trust and Security}
The microservices paradigm poses a number of trust and security challenges. These issues are certainly not new, as they apply to SOA and in general to distributed computing, but they become even more challenging in the context of microservices. In this section, we aim to discuss some of these key security issues.

\paragraph{\textbf{Greater Surface Attack Area}} In monolithic architectures, application processes communicate via internal data structures or internal communication (for instance, socket or RMI). The attack surface is usually also constrained to a single OS. On the contrary, the microservices paradigm is characterised by applications that are broken down into services that interact with each other through APIs exposed to the network. APIs are independent of machine architectures and even programming languages. As a result, they are exposed to more potential attacks than traditional subroutines or functionalities of a large application, which only interacted with other parts of the same application. Moreover, application internals (the microservices) have now become accessible from the external world. Rephrasing, this means that microservices can in principle send the attack surface of a given application through the roof.

\paragraph{\textbf{Network Complexity}} The microservices vision, based on the creation of many small independent applications interacting with each other, can result in complex network activity. This network complexity can significantly increase the difficulty in enforcing the security of the overall microservices-based application. Indeed, when a real-world application is decomposed, it can easily create hundreds of microservices, as seen in the architecture overview of Hailo, an online cab reservation application.\footnote{\url{hailoapp.com}} Such an intrinsic complexity determines an ever-increasing difficulty in debugging, monitoring, auditing, and forensic analysis of the entire application. Attackers could exploit this complexity to launch attacks against applications.

\paragraph{\textbf{Trust}} Microservices, at least in this early stage of development, are often designed to completely trust each other. Considering a microservice trustworthy represents an extremely strong assumption in the ``connectivity era'', where microservices can interact with each other in a heterogeneous and open way. An individual microservice may be attacked and controlled by a malicious adversary, compromising not only the single microservice but, more drastically, bringing down the entire application. As an illustrative real world example, a subdomain of Netflix was recently compromised, and from that domain, an adversary can serve any content in the context of \url{netflix.com}. In addition, since Netflix allowed all users' cookies to be accessed from any subdomain, a malicious individual controlling a subdomain was able to tamper with authenticated Netflix subscribers and their data~\cite{SunNanJae15}. Future microservices platforms need mechanisms to monitor and enforce the connections among microservices to confine the trust placed on individual microservices, limiting the potential damage if any microservice gets compromised.

\paragraph{\textbf{Heterogeneity}} The microservices paradigm brings heterogeneity (of distributed systems) to its maximum expression. Indeed, a microservices-based system can be characterised by: a large number of autonomous entities that are not necessarily known in advance (again, trust issue); a large number of different administrative security domains, creating competition amongst providers of different services; a large number of interactions across different domains (through APIs); no common security infrastructure (different ``Trusted Computing Base’’); and last but not least, no global system to enforce rules.

The research community is still far from adequately addressing the aforementioned security issues. Some recent works, like~\cite{SunNanJae15}, show that some preliminary contribution is taking place.  However, the challenge of building secure and trustworthy microservices-based systems is still more than open.

\section{Conclusions}
\label{sec:conclusions}
The microservice architecture is a style that has been increasingly gaining popularity in the last few years, both in academia and in the industrial world. In particular, the shift towards microservices is a sensitive matter for a number of companies involved in a major refactoring of their back-end systems \cite{DDLM2017}.

Despite the fact that some authors present it from a \emph{revolutionary} 
perspective, we have preferred to provide an \emph{evolutionary} presentation to help the reader understand the main motivations that lead to the distinguishing characteristics of microservices and relate to well-established paradigms such as OO and SOA. With microservice architecture being very recent, we have not found a sufficiently comprehensive collection of literature in the field, so that we felt the need to provide a starting point for newcomers to the discipline,
and offer the authors' viewpoint on the topic.

In this chapter, we have presented a (necessarily incomplete) overview of software architecture, mostly providing the reader with references to the literature, and guiding him/her in our itinerary towards the advent of services and microservices. A specific arc has been given to the narrative, which necessarily emphasises some connections and some literature, and it is possibly too severe with other sources. For example, research contributions in the domain of the actor model \cite{Hewitt:1973} and software agents \cite{KER:Nwana1996} have not been emphasised enough, and still modern distributed systems have been influenced by these communities too. This calls for a broader survey  investigating relationships along this line.
For information on the relation between microservices and scalability, the reader may refer to \cite{dragoni2017}.

\subsection*{Acknowledgements}
Montesi was supported by CRC (Choreographies for Reliable and efficient Communication software), grant no. DFF--4005-00304 from the Danish Council for Independent Research. Giallorenzo was supported by the EU EIT Digital project SMAll. This work has been partially funded by an Erasmus Mundus Scholarship. We would like to thank 
Daniel Martin Johnston who played a major role in proofreading the final draft of the paper and improving the quality of writing.
%
%
\bibliographystyle{plain}
\bibliography{Mazzara_microservices_biblio}

\end{document}